\documentclass[reprint]{revtex4}
\usepackage{amsfonts}
\usepackage{amssymb}
\usepackage{amsmath}
\usepackage{hyperref}
\usepackage{latexsym}
\usepackage[dvips]{graphicx}
\usepackage{epsf}

\begin{document}

\title{Effective action for a free scalar field in the presence of spacetime
foam }
\author{A.A. Kirillov}
\author{ E.P. Savelova}
\email{sep$\_$ 22.12.79@inbox.ru} \affiliation{Dubna State
 University, Universitetskaya Str. 19, Dubna, 141980, Russia }
\date{}

\begin{abstract}
We model spacetime foam by a gas of virtual wormholes. For a free
scalar field we derive the effective Lagrangian which accounts for
the interaction with spacetime foam and contains two additional
non-local terms. One term describes the scattering of scalar
particles on virtual wormholes and explicitly reproduces the
Pauli-Villars regularization procedure. The second term describes
the back reaction of particles on the number density of wormholes
and introduces a self-interaction between particles.
\end{abstract}

\maketitle

\section{Introduction}

At very small scales the spacetime has the foam-like picture \cite%
{wheelerX,H78X} which can be modelled by a gas of virtual wormholes \cite%
{KS10X,KS10Xa,KS10Xb}. We point out that gas of actual
(astrophysical) wormholes was first investigated in a series of
papers \cite{KS,KSa,KSb} where we have demonstrated that actual
wormholes may be responsible for the dark matter phenomenon.
Unlike astrophysical wormholes virtual wormholes exist only for
very small period of time and at very small scales. Such  objects
(virtual wormholes) were first suggested in Refs.
\cite{Loss1,Loss2,Loss3} where it was proposed that they may lead
to loss of quantum coherence. It was latter shown
\cite{Lambda,Lambda2,Lambda3} that quantum coherence is not lost,
since the effects of such wormholes can be absorbed into a
redefinition of coupling
constants of the low energy theory (see also analogous result in Refs. \cite%
{S12ef,S14}). However, they still may play an important role in particle
physics at very high energies, for they may introduce in a natural way the
cutoff at very small scales and may remove divergencies in quantum field
theory \cite{KS10X,KS10Xa,KS10Xb,S12ef}. Moreover, in the presence of
external fields the number density of virtual wormholes changes \cite%
{KS13a,KS14} which gives the principle possibility to form wormhole-like
objects in laboratory. It also may help to explain a non-vanishing small
value of the cosmological constant \cite{KS10X,KS10Xa}.

In Refs \cite{KS13a,KS14} we however considered the case when external
fields do not violate the homogeneity of vacuum state which is too
restrictive. In the present paper we continue our study by inferring the
effective Lagrangian for a free scalar field interacting with the spacetime
foam. As we shall see the effective Lagrangian includes two additional
non-local terms. One term introduces the linear dispersion (which describes
scattering of particles on the foam, i.e., on virtual wormholes). That term
is responsible for the removal of divergencies in the theory and it seems to
reproduce the well-known Pauli-Willars procedure of the regularization \cite%
{reg,reg2}. The second term corresponds to the first nonlinear
correction which describes the change of the wormhole number
density in an arbitrary (inhomogeneous) external field. Such a
term introduces a self-interaction between particles which appears
due to the back reaction of the presence of scalar particles on
the number density of virtual wormholes. We point out that in the
case of a fixed background topology, when the back reaction is
absent, the last non-linear term is always absent, i.e. the scalar
field remains to be free. However in the case of spacetime foam
topology fluctuations change in the presence of scalar particles
and such a non-linear term allows to describe the effect of the
redistribution of virtual wormholes in external classical fields.

\section{Virtual wormhole}

In what follows we shall use some of our previous results \cite%
{KS10X,KS10Xa,S14,KS14}. A virtual wormhole is described as follows.
Consider the metric ($\alpha =1,2,3,4$)%
\begin{equation}
ds^{2}=h^{2}\left( r\right) \delta _{\alpha \beta }dx^{\alpha }dx^{\beta },
\label{X_wmetr}
\end{equation}%
where
\begin{equation}
h\left( r\right) =1+\theta \left( a-r\right) \left( \frac{a^{2}}{r^{2}}%
-1\right),
\end{equation}%
 $\theta \left( x\right) $ is the step function, and
 $r^2=\delta _{\alpha \beta }x^{\alpha }x^{\beta }$. Such a wormhole has
vanishing throat length. Indeed, in the region $r>a$, $h=1$ and
the metric is flat, while the region $r<a$, with the obvious
transformation $y^{\alpha }=\frac{a^{2}}{r^{2}}x^{\alpha }$, is
also flat for $y>a$. Therefore, the regions $r>a$ and $r<a$
represent two Euclidean spaces glued at the surface of a sphere
$S^{3}$ with the center at the origin $r=0$ and radius $r=a$. Such
a space can be described with the ordinary double-valued flat
metric in the region $r_{\pm }>a$ by
\begin{equation}
ds^{2}=\delta _{\alpha \beta }dx_{\pm }^{\alpha }dx_{\pm }^{\beta },
\label{X_wmetr2}
\end{equation}%
where the coordinates $x_{\pm }^{\alpha }$ describe two different sheets of
space. We point out that in the quasi-classical region a virtual wormhole
may be taken as a solution of the Euclidean Einstein equations and the
function $h$ should be smooth. In particular, the choice $%
h(r)=(r^{2}+a^{2})/r^{2}$ corresponds to the so-called
Bronnikov-Ellis metric \cite{br,el}
\begin{equation*}
ds^{2}=dR^{2}+(R^{2}+4a^{2})d\Omega ^{2},
\end{equation*}%
where $d\Omega ^{2}$ is the angular part of the metric, $R=r-a^{2}/r$ and $%
-\infty <R<\infty $. There also exist classes of wormhole metrics
which possess less symmetry (e.g., for cylindrically and axially
symmetric wormholes see \cite{br2} and references therein).
 However as it
was shown earlier in \cite{KS10X,KS10Xa} for wormholes with the
characteristic size of throats $a\ll \ell_{Pl}$ the contribution
to the action from the curvature is negligible as compared to the
zero-point fluctuations and therefore the model metric
(\ref{X_wmetr2}) is sufficient to our aims. Moreover, in the
complete quantum gravity the partition function (which for the
scalar field is calculated latter) assumes the sum over all field
configurations and, therefore, it is not important which kind of
the background metric is used.

Let identify the inner and outer regions of the sphere $S^{3}$ and construct
a wormhole which connects regions in the same space (instead of two
independent spaces). This is achieved by gluing the two spaces in (\ref%
{X_wmetr2}) by motions of the Euclidean space (the Poincare motions). If $%
R_{\pm }$ is the position of the sphere in coordinates $x_{\pm }^{\mu }$,
then the gluing is the rule%
\begin{equation}
x_{+}^{\mu }=R_{+}^{\mu }+\Lambda _{\nu }^{\mu }\left( x_{-}^{\nu
}-R_{-}^{\nu }\right) ,  \label{X_gl}
\end{equation}%
where $\Lambda _{\nu }^{\mu }\in O(4)$, which represents the composition of
a translation and a rotation of the Euclidean space. In terms of common
coordinates such a wormhole represents the standard flat space in which the
two spheres $S_{\pm }^{3}$ (with centers at positions $R_{\pm }$) are glued
by the rule (\ref{X_gl}). We point out that the physical region is the outer
region of the two spheres. Thus, in general, the wormhole is described by a
set of parameters $\xi $: the throat radius $a$, positions of throats $%
R_{\pm }$ , and rotation matrix $\Lambda _{\nu }^{\mu }\in O(4)$.

\section{Green function in a gas of virtual wormholes}

The Green function in a gas of virtual wormholes was considered
first in  \cite{S12ef}. It was shown there that the exact Green
function admits the representation in the form of an infinite
series and, therefore, it has too complicated form. In the present
paper our primary aim is to investigate the structure and
properties of the effective action and, therefore, we will
restrict to the simplest case of the minimally coupled scalar
field.

 Consider
now the simplest scalar field  and construct the Green function in
the presence of a gas of wormholes. The Green function obeys the
Laplace equation
\begin{equation*}
\left( -\Delta +m^{2}\right) G\left( x,x^{\prime }\right) =\delta \left(
x-x^{\prime }\right)
\end{equation*}%
with proper boundary conditions at throats (we require $G$ and the
normal derivative $\partial G/\partial n$ to be continuous at
throats, where $n$ is the unit vector normal to the surface of the
throat section). The Green function for the Euclidean space is
merely
\begin{equation}
G_{0}(x-x^{\prime })=\frac{m^{2}}{4\pi ^{2}}\frac{K_{1}\left( mr\right) }{mr}
\label{grf0}
\end{equation}%
where $r^{2}=\left( x-x^{\prime }\right) ^{2}$(and $G_{0}\left( k\right)
=1/(k^{2}+m^{2})$ for the Fourier transform). In the massless case the Green
function reduces to $G_{0}(x,x^{\prime })=1/(4\pi ^{2}\left( x-x^{\prime
}\right) ^{2})$.

In the presence of a single wormhole which connects two Euclidean spaces
this equation admits the exact solution. For the outer region of the throat $%
S^{3}$ the source $\delta \left( x-x^{\prime }\right) $ generates
a set of 4-dimensional multipoles placed in the center of sphere
which gives the corrections to the Green function $G_{0}+\delta
G$. In the present paper we restrict to the lowest monopole term
only (see more details in \cite{S12ef,S14}) which gives
\begin{equation}
\delta G_{\pm }=\mp 2\pi ^{2}a^{2}G_{0}(r^{\prime })G_{0}(r_{\pm }),
\end{equation}%
where $\delta G_{+}$ corresponds to $r>a$ and $r_{+}=r$, while $\delta G_{-}$
corresponds to the region $r<a$ and $r_{-}=a^{2}/r$. We also assume here $%
ma\ll 1$ since virtual wormholes are expected to have the Planckian size. A
single wormhole which connects two regions in the same space is a couple of
conjugated spheres $S_{\pm }^{3}$ of the radius $a$ with a distance $\vec{X}=%
\vec{R}_{+}-\vec{R}_{-}$ between centers of spheres. So the parameters of
the wormhole are $\xi =(a,R_{+},R_{-})$. The interior of the spheres is
removed and surfaces are glued together. Then in the approximation $a/X\ll 1$
the correction to the Green function can be taken as

\begin{equation}
\delta G(x,x^{\prime })=-2\pi ^{2}a^{2}\left( G_{0}\left( x_{+}\right)
-G_{0}\left( x_{-}\right) \right) \left( G_{0}\left( x_{+}^{\prime }\right)
-G_{0}\left( x_{-}^{\prime }\right) \right) ,  \label{gr0}
\end{equation}%
where we denote $x_{\pm }=x-R_{\pm }$. The above expression explicitly shows
the symmetry $x\longleftrightarrow x^{\prime }$. When we consider a dilute
gas approximation the correction to the Green function becomes additive and
can be written as
\begin{equation}
\delta G(x,x^{\prime })=\sum \delta G\left( x,x^{\prime },\xi _{i}\right)
=\int \delta G\left( x,x^{\prime },\xi \right) F(\xi )d\xi ,  \label{gr}
\end{equation}%
where
\begin{equation}
F\left( \xi \right) =\sum\limits_{i}\delta \left( \xi -\xi _{i}\right)
\label{X_F}
\end{equation}%
is the density of wormholes in the configuration space $\xi $. In the vacuum
case the background distribution has an isotropic and homogeneous character,
i.e., $\rho (\xi )=<0|F\left( \xi \right) |0>$ with $\rho (\xi )=\rho \left(
a,X\right) $, then for the Fourier transforms of the Green function we find%
\begin{equation}
G\left( k\right) =\frac{1}{k^{2}+m^{2}}\left( 1-\frac{1}{k^{2}+m^{2}}\nu
\left( k\right) \right) \approx \frac{1}{k^{2}+m^{2}+\nu \left( k\right) },
\label{grb}
\end{equation}%
where%
\begin{equation}
\nu \left( k\right) =4\pi ^{2}\int a^{2}\left( \rho \left( a,0\right) -\rho
\left( a,k\right) \right) da,  \label{X_bk}
\end{equation}%
and $\rho \left( a,X\right) =\int \rho \left( a,k\right) e^{-ikX}\frac{d^{4}k%
}{\left( 2\pi \right) ^{4}}$.

\section{Generating functional}

Consider now the generating functional (the partition function) which is
used to generate all possible correlation functions in quantum field theory
(and the perturbation scheme when we include interactions)
\begin{equation}
Z_{total}\left( J\right) =\sum\limits_{\tau }\sum\limits_{\varphi }e^{-S_{E}}
\end{equation}%
where the sum is taken over field configurations $\varphi $ and topologies $%
\tau $ (wormholes). The Euclidean action is
\begin{equation}
S_{E}=\frac{1}{2}\left( \varphi \left( -\Delta +m^{2}\right) \varphi \right)
-\left( J\varphi \right) ,  \label{X_act}
\end{equation}%
and we use the notions
\begin{equation*}
\left( J\varphi \right) =\int J\left( x\right) \varphi \left( x\right)
d^{4}x.
\end{equation*}%
Here $J$ denotes an external current. The sum over field configurations $%
\varphi $ can be replaced by the integral
\begin{equation}
Z^{\ast }\left( J\right) =\int \left[ D\varphi \right] e^{\frac{1}{2}\left(
\varphi \left( \Delta -m^{2}\right) \varphi \right) +\left( J\varphi \right)
}.  \label{X_gf1}
\end{equation}%
Upon the simple transformations
\begin{equation}
\frac{1}{2}\left( \varphi \left( -\Delta +m^{2}\right) \varphi \right)
-\left( J\varphi \right) =\frac{1}{2}\left( \widetilde{\varphi }(-\Delta
+m^{2})\widetilde{\varphi }\right) -\frac{1}{2}\left( JGJ\right) ,
\end{equation}%
where $\widetilde{\varphi }=\varphi -GJ$ and $G~$is the background Green
function $G=G_{0}+\delta G(\xi )$, e.g., see (\ref{grf0}) and (\ref{gr}), we
cast the partition function to the form
\begin{equation}
Z^{\ast }=\int \left[ D\widetilde{\varphi }\right] e^{-\frac{1}{2}\left(
\widetilde{\varphi }(-\Delta +m^{2})\widetilde{\varphi }\right) +\frac{1}{2}%
\left( JGJ\right) }=Z_{0}(G)e^{\frac{1}{2}\left( JGJ\right) },  \label{X_gf2}
\end{equation}%
where $Z_{0}(G)=\int \left[ D\varphi \right] e^{-\frac{1}{2}\left( \varphi
(-\Delta +m^{2})\varphi \right) }$ is the standard expression and $G=G\left(
\xi _{1},...,\xi _{N}\right) $ is the Green function for a fixed topology,
i.e., for a fixed set of wormholes $\xi _{1},...,\xi _{N}$ .

Consider now the sum over topologies $\tau $. To this end we restrict with
the sum over the number of wormholes and integrals over parameters of
wormholes:
\begin{equation}
\sum\limits_{\tau }\rightarrow \sum\limits_{N}\int
\prod\limits_{i=1}^{N}d\xi _{i}=\int \left[ DF\right]  \label{X_ts}
\end{equation}%
where $F$ is given by (\ref{X_F}). We point out that in general the
integration over parameters is not free (e.g., it obeys the obvious
restriction $\left\vert \vec{R}_{i}^{+}-\vec{R}_{i}^{-}\right\vert \geq
2a_{i}$). This defines the generating function as
\begin{equation}
Z_{total}\left( J\right) =\int \left[ DF\right] Z_{0}(G)e^{\frac{1}{2}\left(
JGJ\right) }.
\end{equation}%
Expanding this expression by $J$ we find
\begin{equation}
W\left( J\right) -W(0)=\ln \frac{Z_{tot}\left( J\right) }{Z_{tot}\left(
0\right) }\approx \frac{1}{2}\overline{\left( JGJ\right) }+\frac{1}{8}%
\overline{\left( J\Delta GJ\right) ^{2}}+\frac{1}{48}\overline{\left(
J\Delta GJ\right) ^{3}}+...  \label{W}
\end{equation}%
where overbar denotes vacuum mean value $\overline{G}=<0|G|0>$, i.e.,
\begin{equation*}
\overline{G}=<0|G|0>_{J=0}=\frac{1}{Z_{total}\left( 0\right) }\int \left[ DF%
\right] Z_{0}(G)G\left( \xi \right) .
\end{equation*}%
The two terms in (\ref{W}) can be expressed via moments of the density of
wormholes in the configuration space as follows%
\begin{equation*}
\overline{G}=G_{0}+\int \delta G\left( \xi \right) \rho (\xi )d\xi ,
\end{equation*}%
where
\begin{equation*}
\rho (\xi )=<0|F\left( \xi \right) |0>_{J=0}
\end{equation*}%
is the mean density (\ref{X_F}) and $\delta G\left( \xi \right) $ is given
by (\ref{gr0}). Analogously the next term in (\ref{W}) which describes
topology fluctuations can be expressed as%
\begin{equation*}
\overline{\Delta G\Delta G}=\int \delta G\left( \xi \right) \delta G\left(
\xi \right) \rho (\xi )d\xi +\int \delta G\left( \xi \right) \delta G\left(
\xi ^{\prime }\right) \omega (\xi ,\xi ^{\prime })d\xi d\xi ^{\prime }
\end{equation*}%
where we denote ($\Delta F\left( \xi \right) =F\left( \xi \right) -\rho (\xi
)$)%
\begin{equation*}
\rho (\xi ,\xi ^{\prime })=<0|\Delta F\left( \xi \right) \Delta F\left( \xi
^{\prime }\right) |0>_{J=0}=\rho (\xi )\delta \left( \xi -\xi ^{\prime
}\right) +\omega (\xi ,\xi ^{\prime }).
\end{equation*}%
All higher order mean values in (\ref{W}), e.g., $\overline{\Delta G\Delta
G\Delta G}$, are expressed via the respective momenta $\rho (\xi ,\xi
^{\prime },\xi ^{\prime \prime })$, etc. as
\begin{equation*}
\overline{\Delta G\Delta G\Delta G}=\int \delta G\left( \xi \right) \delta
G\left( \xi ^{\prime }\right) \delta G\left( \xi ^{\prime \prime }\right)
\rho (\xi ,\xi ^{\prime },\xi ^{\prime \prime })d\xi d\xi ^{\prime }d\xi
^{\prime \prime }.
\end{equation*}%
We use the approximation of a rarefied gas of virtual wormholes and,
therefore, to the leading order we may assume that the correlations between
positions of wormholes are absent, i.e., $\omega (\xi ,\xi ^{\prime
})\approx 0$. To account for such correlations it requires the further
development of the theory. Thus, we find the decomposition
\begin{equation}
W\left( J\right) -W_{0}=\frac{1}{2}\left( JG_{0}J\right) +\frac{1}{2}\int
\left( J\delta G\left( \xi \right) J\right) \rho (\xi )d\xi +\frac{1}{8}\int
\left( J\delta G\left( \xi \right) J\right) ^{2}\rho (\xi )d\xi +...
\end{equation}%
The first term here corresponds to the standard free scalar field. The
second term describes effects of the scattering of scalar particles on the
non-trivial topology. We point out that if the topology does not change by
the presence of scalar particles (i.e., it is rigidly fixed, though it may
be random) next terms do not appear at all. While in general there appears a
non-linearity (due to the back reaction). In other words, in the presence of
spacetime foam the scalar field inevitably acquires a self-interaction (the
last term above) which means a modification of the field theory. The more
convenient way to analyze such a modification is to consider the effective
action.

\section{Effective action}

Consider the mean vacuum value of the scalar field which is given by $%
\varphi _{c}(J)=<0|\varphi |0>_{J}=\frac{\delta W}{\delta J}$ or
\begin{equation*}
\varphi _{c}(J)=G_{0}J+\int \delta G\left( \xi \right) J\rho (\xi )d\xi +%
\frac{1}{2}\int \left( J\delta G\left( \xi \right) J\right) \delta G\left(
\xi \right) J\rho (\xi )d\xi +...
\end{equation*}%
This equation can be resolved $J(\varphi _{c})$, then we define the
effective action
\begin{equation*}
\Gamma (\varphi _{c})=(\varphi _{c}J(\varphi _{c}))-W(J(\varphi _{c})).
\end{equation*}%
This transformation can be carried out by the perturbation method. Indeed, $%
\delta G\left( \xi \right) $ includes a small parameter $a^{2}$ which is
expected to have the order $a\sim \ell _{Pl}$. Thus we find
\begin{equation*}
\Gamma (\varphi )=\frac{1}{2}(\varphi \left( -\Delta +m^{2}\right) \varphi
)+V_{1}(\varphi )+V_{2}(\varphi )+...
\end{equation*}%
where
\begin{equation*}
V_{1}(\varphi )=-\frac{1}{2}\int \varphi \left( -\Delta +m^{2}\right) \delta
G\left( \xi \right) \left( -\Delta +m^{2}\right) \varphi \rho (\xi )d\xi
\end{equation*}%
and
\begin{equation*}
V_{2}(\varphi )=-\frac{1}{8}\int \left( \varphi \left( -\Delta +m^{2}\right)
\delta G\left( \xi \right) \left( -\Delta +m^{2}\right) \varphi \right)
^{2}\rho (\xi )d\xi .
\end{equation*}%
In the explicit form (by means of use (\ref{gr0})) we find
\begin{equation*}
V_{1}(\varphi )=\pi ^{2}\int a^{2}\left( \varphi \left( R_{+}\right)
-\varphi \left( R_{-}\right) \right) ^{2}\rho (\xi )d\xi
\end{equation*}%
and
\begin{equation*}
V_{2}(\varphi )=-\frac{\pi ^{4}}{2}\int a^{4}\left( \varphi \left(
R_{+}\right) -\varphi \left( R_{-}\right) \right) ^{4}\rho (\xi )d\xi .
\end{equation*}
We point out that this term is negative which reflects the well-known fact
that the second order correction to the ground state is always negative.

Next terms describe higher order corrections, e.g., the multipole
contribution in (\ref{gr0}) and correlations between positions of wormholes,
e.g.,
\begin{equation*}
V_{3}(\varphi )=-\frac{\pi ^{4}}{2}\int a^{2}\left( \varphi _{+}-\varphi
_{-}\right) ^{2}a^{\prime 2}\left( \varphi _{+}^{\prime }-\varphi
_{-}^{\prime }\right) ^{2}\omega (\xi ,\xi ^{\prime })d\xi d\xi ^{\prime }.
\end{equation*}

Now we are ready to analyze the effective potential $U=V_{1}+V_{2}$. \ The
first term $V_{1}$ defines the renormalization of the Green function in the
form (\ref{X_bk}) which reflects the scattering of scalar particles on
virtual wormholes. Nevertheless for the low energy theory it is convenient
to expand this potential in series. Indeed, the definition of $U$ includes
the unknown function $\rho (\xi )$ which requires the further studying. In
the vacuum state the space remains to be isotropic and homogeneous.
Therefore, we may expect that $\rho (\xi )$ has the structure $\rho (\xi
)=\rho \left( a,\left\vert X\right\vert \right) $, where $X=R_{+}-R_{-}$,
and $V_{1}$ can be rewritten as
\begin{equation*}
V_{1}(\varphi )=\pi ^{2}\int a^{2}\left( \varphi \left( x+X\right) -\varphi
\left( x\right) \right) ^{2}\rho (a,X)dad^{4}Xd^{4}x.
\end{equation*}%
This part explicitly reduces to the correction (\ref{X_bk}). Nevertheless it
has sense to consider its decomposition. For the spacetime foam we expect
the typical values $X,a\sim \ell _{Pl}$ and this allows to expand $V_{1}$ by
the small parameter $X$ as
\begin{equation*}
V_{1}(\varphi )= \pi ^{2}\int a^{2}\left( \varphi \left( x\right) \left(
1-e^{-X^{\mu }\nabla _{\mu }}+1-e^{X^{\mu }\nabla _{\mu }}\right) \varphi
\left( x\right) \right) \rho (a,X)dad^{4}Xd^{4}x,
\end{equation*}%
\begin{equation*}
V_{1}(\varphi )= -2\pi ^{2}\sum_{n=1}^{\infty }\int a^{2}\varphi \left(
x\right) \left( \frac{X^{2n}}{\left( 2n\right) !}\left( s^{\mu }\nabla _{\mu
}\right) ^{2n}\right) \varphi \left( x\right) \rho (a,X)dad^{4}Xd^{4}x,
\end{equation*}%
where $s^{\mu }=X^{\mu }/X$, which gives%
\begin{equation}
V_{1}(\varphi )\approx -\frac{1}{2}\int \varphi \left( A\Delta +\frac{1}{%
M_{1}^{2}}\Delta ^{2}+..\right) \varphi d^{4}x.  \label{V1}
\end{equation}%
Here coefficients are
\begin{equation*}
A=\frac{\pi ^{2}}{2}\int a^{2}X^{2}\rho (a,X)dad^{4}X=\frac{\pi ^{2}}{2}%
<a^{2}X^{2}>n
\end{equation*}%
\begin{equation*}
\frac{1}{M_{1}^{2}}=\frac{\pi ^{2}}{16}\int a^{2}X^{4}\rho (a,X)dad^{4}X=%
\frac{\pi ^{2}}{16}<a^{2}X^{4}>n
\end{equation*}%
where $n=$ $\int \rho (a,X)dad^{4}X$ is the density of wormholes and we have
used properties of the isotropy of the vacuum distribution $\rho (\xi )$,
i.e.,
\begin{equation*}
<s^{\alpha }s^{\beta }>=\frac{1}{4}\delta ^{\alpha \beta },
\end{equation*}%
\begin{equation*}
<s^{\alpha }s^{\beta }s^{\mu }s^{\nu }>=\frac{1}{24}\left( \delta ^{\alpha
\beta }\delta ^{\mu }{}^{\nu }+\delta ^{\alpha \mu }\delta {}^{\nu \beta
}+\delta ^{\alpha \nu }\delta {}^{\beta \mu }\right) .
\end{equation*}%
Rigorously speaking the second term in (\ref{V1}) should also include
contribution from 4-dimensional dipole component in (\ref{gr0}). However the
symmetry of the vacuum means that such a term has the same structure with a
coefficient $1/M^{2}\sim <a^{4}X^{2}>n$. It is clear that such a
contribution can be absorbed in the coefficient $M_{1}$.

In the same way we find the decomposition of the self-interacting term as%
\begin{equation*}
V_{2}(\varphi )=-\frac{1}{4!}\int \frac{1}{M_{2}^{4}}\left( \nabla \varphi
\right) ^{4}d^{4}x+...
\end{equation*}%
where
\begin{equation*}
\frac{1}{M_{2}^{4}}=\frac{\pi ^{4}}{6}<a^{4}X^{4}>n.
\end{equation*}%
Thus for the effective Lagrangian we get
\begin{equation}
\Gamma (\varphi )=\int \left[ \frac{1}{2}\varphi \left( -\Delta (1+A+\frac{1%
}{M_{1}^{2}}\Delta )+m^{2}\right) \varphi -\frac{1}{4!M_{2}^{4}}\left(
\nabla \varphi \right) ^{4}\right] d^{4}x+...  \label{efact}
\end{equation}
We point out that the specification of all possible momenta (i.e., of all
constants in the expression above) is simply an equivalent way to the
definition of the vacuum distribution of virtual wormholes $\rho (\xi )$.
Besides, since this expansion involves the infinite series of constants, the
resulting theory belongs formally to the class of non-remormalizable
theories (as quantum gravity is).

\section{Discussions}

Consider first the quadratic part $\Gamma _{0}(\varphi )$ of the effective
action (\ref{efact}). The dimensionless parameter $A$ merely renormalizes
the naked values of $m$ and $\varphi $. Therefore, this constant can be
excluded, e.g., we may take $A=0$. The naive estimate for the rest
parameters is $M_{a}\sim M_{Pl}\gg m$. In particular, Abdo et al. \cite%
{LorentzCorr} have recently found a very stringent bound $l_{1}<L_{Pl}$,
where $L_{Pl}$ is the Planck length, on the first nontrivial correction to
the standard photon dispersion relation $\omega
^{2}=k^{2}(1+kl_{1}+k^{2}l_{2}^{2}+\ldots )$. The isotropy and homogeneity
of vacuum forbid corrections of the type $kl_{1}$ which means that the first
correction is quadratic. Nevertheless, we can accept such an estimate, then
the effective action can be recast into the form%
\begin{equation}
\Gamma _{0}(\varphi )\approx \frac{1}{2}\int \left( k^{2}+m^{2}-\frac{1}{%
M_{1}^{2}}k^{4}+\frac{1}{M_{3}^{4}}k^{6}+...\right) \left\vert \varphi
_{k}\right\vert ^{2}\frac{d^{4}k}{\left( 2\pi \right) ^{4}}.  \label{gg}
\end{equation}%
If we neglect all terms higher than $k^{4}$ then we get%
\begin{equation*}
\Gamma _{0}(\varphi )\approx \frac{1}{2}\int \left( \left( k^{2}+\widetilde{m%
}^{2}\right) \frac{1}{-M_{1}^{2}}\left( k^{2}+\widetilde{M}_{1}^{2}\right)
\right) \left\vert \varphi _{k}\right\vert ^{2}\frac{d^{4}k}{\left( 2\pi
\right) ^{4}}+...
\end{equation*}%
where $\widetilde{M}_{1}^{2}=-M_{1}^{2}(\sqrt{1+4m^{2}/M_{1}^{2}}+1)/2$ $%
\approx -M_{1}^{2}$ and $\widetilde{m}^{2}=M_{1}^{2}(\sqrt{1+4m^{2}/M_{1}^{2}%
}-1)/2$ $\approx m^{2}$. Let us forget for a while that one mass has the
negative sign $\widetilde{M}_{1}^{2}$ $\approx -M_{1}^{2}$. We see that at
sufficiently small energies $k^{2}\ll M_{1}^{2}$ the above action describes
the standard free scalar particles and effects of the spacetime foam are
negligible (they are merely encoded in the renormalized parameters). While
at Planck energies (due to the interaction with the foam) there appear in
the theory new (additional) particles with a very huge mass $\widetilde{M}%
_{1}\sim M_{Pl}$. From the formal standpoint this part recovers
the well-known Feynman or Pauli-Villars regularization procedure
\cite{reg,reg2}. It becomes also clear the general structure of
the above action which can also be seen from (\ref{X_bk}) (e.g.,
$\nu \left( k\right) =\sum \left( \pm
L_{n}^{2}k^{2}\right) ^{n}$). Next terms of the decomposition of $%
V_{1}(\varphi )$ will add new additional and more massive terms which
generate additional degrees of freedom, e.g.,
\begin{equation}
\Gamma _{0}(\varphi )=\frac{1}{2}\int \varphi \left( -\Delta +\widetilde{m}%
^{2}\right) \prod\limits_{a}\left( \frac{-\Delta +\widetilde{M}_{a}^{2}}{%
\widetilde{M}_{a}^{2}}\right) \varphi d^{4}x  \label{PW}
\end{equation}%
where $\widetilde{m}$, $\widetilde{M}_{a}$ are the physical (i.e., already
renormalized) values of the mass spectrum for the scalar particles in the
theory.

This action contains a number of extra degrees of freedom which
are very heavy particles with masses $\widetilde{M}_{a}\gtrsim
M_{Pl}$. Such additional particles relate to the modes defined on
the wormhole necks and more complex collective excitations.
Indeed, if we consider a constant time sections then, in general,
such a section splits into the Euclidean 3-dimensional space and a
number of closed spaces (throats or baby universes). Since all
such throats have the Planckian size then the respective modes
have wavelengths which start from Planckian values. This explains
why such additional particles are extremely heavy.

The only annoying thing with the above action is that some part of
spectrum has imaginary (or imaginary parts) masses, e.g.,
$\widetilde{M}_{1}^{2}$ $\approx $ $-M_{1}^{2}$ $<$ $0$. For
particle physics (upon continuation to the Minkowsky sector) this
means the presence of instabilities, the energy of particles
$\varepsilon =\sqrt{p^{2}-M_{1}^{2}}$ is imaginary as $p<M_{1}$.
Such an instability represents the well-known and essential
property of wormholes which corresponds to the negative mode found
first in Refs. \cite{R,R2}. It describes the instability of a
large universe against the emission of baby universes and may
serve as one of the simplest examples of theories which admit
solutions violating the null energy condition \cite{R3} (see also
references therein).

 The stable vacuum corresponds to the case when
all masses have values $\widetilde{M}_{a}^{2}>0$. Frankly speaking
we do not know what kind of Planck-scale physics is hidden here
and may only suggest some speculations. The fist and simplest
possibility is that higher order corrections should include
multipoles of higher orders in (\ref{gr0}) which may properly
change the values of the mass spectrum. This however is too
optimistic point of view. Indeed, baby universes correspond to the
virtual wormholes with sufficiently long necks. Then the monopole
component is absent and the first contribution in (\ref{gr0})
comes from the dipole. Nevertheless, the instability retains as it
was shown in \cite{R,R2}. The second possibility is that the
restriction to a finite number of terms in the above action is
simply incorrect. Indeed, if we take, as an example, a particular form of $%
\rho (\xi )$, e.g., see \cite{S14,KS14}
\begin{equation*}
\rho (\xi )=n\delta \left( a-a_{0}\right) \frac{1}{2}\left( \delta
^{4}\left( X-r_{0}\right) +\delta ^{4}\left( X+r_{0}\right) \right) ,
\end{equation*}%
we find $\nu \left( k\right) =4\pi ^{2}na_{0}^{2}\left( 1-\cos \left(
kr_{0}\right) \right) \geq 0$. Then in the general case when $%
n=n(a_{0},r_{0})$ we get%
\begin{equation*}
\nu \left( k\right) =\int n(a,X)a^{2}\left( 1-\cos \left( kX\right) \right)
dadX\geq 0
\end{equation*}%
which means that the quadratic part of the action
\begin{equation*}
\Gamma _{0}(\varphi )=\frac{1}{2}\int \left( k^{2}+m^{2}+\nu \left( k\right)
\right) \left\vert \varphi _{k}\right\vert ^{2}\frac{d^{4}k}{\left( 2\pi
\right) ^{4}}
\end{equation*}%
actually contains no poles on the real axis $k$ (at least in the Euclidean
sector). Though, formally the expansion of $\cos \left( rk\right) $ gives
the series of the alternating sign (exactly as in (\ref{gg})). This however
does not guarantee the absence of poles on the complex plane $k$ and the
stability of the vacuum (the stable vacuum contains only poles of the type $%
k_{a}=\pm i\widetilde{M}_{a}$). Here we come to the third
possibility that the instabilities have the direct physical sense
and lead to baby universes production \cite{R,R2}. Such an
instability should be accompanied with the phase transitions.
Fortunately we have the additional self-interacting terms, e.g.,
$V_{2}$. The instability results in the particle (as well as
actual wormholes or baby universes) production and the change
(redefinition) of the vacuum state (which is described by the
Bogolubov's transformations for an appropriate particles $
\widetilde{M}_{a}$). In other words such an instability (and
actual wormhole production) works till the moment when all masses
get into the stable physical sector $Im\widetilde{M}_{a}=0$. Then
the quadratic part of the action transforms exactly to the
Pauli-Villars type (\ref{PW}).

Consider now the second self-interacting term and continuation to the
Minkowsky space $S=i\Gamma _{0}(\varphi )$, $\left( \nabla \varphi \right)
^{2}\rightarrow -\left( \partial _{\mu }\varphi \right) ^{2}$, $-\Delta
\rightarrow \left( \partial _{\mu }\right) ^{2}$, etc. Then in the low
energy limit we find (we neglect the higher order corrections in the
quadratic part)%
\begin{equation*}
S(\varphi )=\int \left[ \frac{1}{2}\left( \left( \partial _{\mu }\varphi
\right) ^{2}-m^{2}\varphi ^{2}\right) +\frac{1}{4!M_{2}^{4}}\left( -\left(
\partial _{\mu }\varphi \right) ^{2}\right) ^{2}\right] \sqrt{-g}d^{4}x
\end{equation*}%
where $\left( \partial _{\mu }\varphi \right) ^{2}=g^{\mu \nu }\left(
\partial _{\mu }\varphi \right) \left( \partial _{\nu }\varphi \right) $,
etc. The non-linear term here describes the redistribution of virtual
wormholes (the change in the wormhole number density) in an external field.
We point out that such a term does not presumes the homogeneity of the
external field as in Refs. \cite{KS13a,KS14}.

Consider now the presence of an external field $\varphi _{0}$ which should
be a solution to the equations of motions which follow from the above
action, i.e., $\delta S(\varphi _{0})=0$. Then we find the quadratic part of
the action in the presence of the external field $\varphi _{0}$ as
\begin{equation*}
S_{2}(\varphi _{0},\varphi )=\frac{1}{2}\int \left[ \left( \Gamma ^{\mu \nu
}(\varphi _{0})\partial _{\mu }\varphi \partial _{\nu }\varphi -m^{2}\varphi
^{2}\right) \right] \sqrt{-g}d^{4}x
\end{equation*}%
where%
\begin{equation*}
\Gamma ^{\mu \nu }(\varphi _{0})=\left( 1+\frac{1}{3!M_{2}^{4}}n_{\alpha
}n^{\alpha }\right) g^{\mu \nu }+\frac{2}{3!M_{2}^{4}}n^{\nu }n^{\mu }
\end{equation*}%
and $n_{\mu }=\partial _{\mu }\varphi _{0}$. Let the external field has the
property $\partial _{\mu }\varphi _{0}=n_{\mu }=const$. Then $\Gamma ^{\mu
\nu }(\varphi _{0})=const$ and from the above action we find the dispersion
relations in the form (compare to the consideration in Ref. \cite{S14})%
\begin{equation*}
\Gamma ^{\mu \nu }(\varphi _{0})k_{\nu }k_{\mu }-m^{2}=0
\end{equation*}%
Consider the time-like vector $n^{\mu }=(b,0)$ and $k_{\mu }=(\varepsilon ,%
\mathbf{p})$. Then \ we find $n_{\alpha }n^{\alpha }=b^{2}$ and this gives
(we assume $\frac{b^{2}}{3!M_{2}^{4}}\ll 1$)
\begin{equation*}
\varepsilon ^{2}(\mathbf{p})\approx \left( 1-\frac{3b^{2}}{3!M_{2}^{4}}%
\right) m^{2}+\left( 1-\frac{2b^{2}}{3!M_{2}^{4}}\right) \mathbf{p}^{2}.
\end{equation*}%
We see that in this case the speed of light does not exceed the vacuum value
\begin{equation*}
c^{2}=1-\frac{2b^{2}}{3!M_{2}^{4}}<1
\end{equation*}%
which, according to  \cite{S14}, means that the additional density of
virtual wormholes is $\delta n>0$.

Consider now the space-like vector $n^{\mu }=(0,\mathbf{b})$ which gives $%
n_{\alpha }n^{\alpha }=-\mathbf{b}^{2}$ and therefore the dispersion
relations takes the form
\begin{equation*}
\varepsilon ^{2}(\mathbf{p})\approx m^{2}\left( 1-\frac{\mathbf{b}^{2}}{%
3!M_{2}^{4}}\right) +\mathbf{p}_{\bot }^{2}+(1-\frac{3\mathbf{b}^{2}}{%
3!M_{2}^{4}})\mathbf{p}_{\Vert }^{2}
\end{equation*}%
which gives $c_{\bot }^{2}=1$ and $c_{\Vert }^{2}=1-\frac{\mathbf{b}^{2}}{%
2M_{2}^{4}}<1$. According to \cite{S14} this means $\delta n<0$.
In other words the external field may either increase, or suppress
the density of virtual wormholes. There remains a much more
complex situation when the external field retains the dependence
on coordinates which however requires the further investigations.

\end{document}